\documentclass[%
 reprint,
superscriptaddress,
 amsmath,amssymb,
 aps,
 prd,
]{revtex4-2}
\usepackage[left=1in,right=1in,top=1in,bottom=1in]{geometry}
\usepackage[%
  colorlinks=true,
  urlcolor=blue,
  linkcolor=blue,
  citecolor=blue
]{hyperref}
\usepackage{cancel}
\usepackage{graphicx,sidecap}
\usepackage{dcolumn}
\usepackage{bm}
\usepackage{printlen} 
\usepackage{siunitx}
\usepackage{aas_macros}
\DeclareSIUnit \parsec {pc }
\DeclareSIUnit \year {yr}

\newcommand{\revCL}[1]{{#1}} 
\newcommand{\commentMD}[1]{}

\usepackage[mathlines]{lineno}

\date{\today}

\begin{document}

\preprint{APS/123-QED}

\title{Constraining Primordial Black Holes using Fast Radio Burst Gravitational-Lens Interferometry with CHIME/FRB}

\author{Calvin Leung}
  \thanks{These authors contributed equally to this work.}
  \affiliation{MIT Kavli Institute for Astrophysics and Space Research, Massachusetts Institute of Technology, 77 Massachusetts Ave, Cambridge, MA 02139, USA}
  \affiliation{Department of Physics, Massachusetts Institute of Technology, 77 Massachusetts Ave, Cambridge, MA 02139, USA}
\author{Zarif Kader}
  \thanks{These authors contributed equally to this work.}
  \affiliation{Department of Physics, McGill University, 3600 rue University, Montr\'eal, QC H3A 2T8, Canada}
  \affiliation{McGill Space Institute, McGill University, 3550 rue University, Montr\'eal, QC H3A 2A7, Canada}
\author{Kiyoshi W.~Masui}
  \affiliation{MIT Kavli Institute for Astrophysics and Space Research, Massachusetts Institute of Technology, 77 Massachusetts Ave, Cambridge, MA 02139, USA}
  \affiliation{Department of Physics, Massachusetts Institute of Technology, 77 Massachusetts Ave, Cambridge, MA 02139, USA}
\author{Matt Dobbs}
  \affiliation{Department of Physics, McGill University, 3600 rue University, Montr\'eal, QC H3A 2T8, Canada}
  \affiliation{McGill Space Institute, McGill University, 3550 rue University, Montr\'eal, QC H3A 2A7, Canada}
\author{Daniele Michilli}
  \affiliation{MIT Kavli Institute for Astrophysics and Space Research, Massachusetts Institute of Technology, 77 Massachusetts Ave, Cambridge, MA 02139, USA}
  \affiliation{Department of Physics, Massachusetts Institute of Technology, 77 Massachusetts Ave, Cambridge, MA 02139, USA}
\author{Juan Mena-Parra}
  \affiliation{MIT Kavli Institute for Astrophysics and Space Research, Massachusetts Institute of Technology, 77 Massachusetts Ave, Cambridge, MA 02139, USA}
\author{Ryan Mckinven}
  \affiliation{Department of Physics, McGill University, 3600 rue University, Montr\'eal, QC H3A 2T8, Canada}
  \affiliation{McGill Space Institute, McGill University, 3550 rue University, Montr\'eal, QC H3A 2A7, Canada}
\author{Cherry Ng}
  \affiliation{Dunlap Institute for Astronomy \& Astrophysics, University of Toronto, 50 St.~George Street, Toronto, ON M5S 3H4, Canada}
\author{Kevin Bandura}
  \affiliation{Lane Department of Computer Science and Electrical Engineering, 1220 Evansdale Drive, PO Box 6109, Morgantown, WV 26506, USA}
  \affiliation{Center for Gravitational Waves and Cosmology, West Virginia University, Chestnut Ridge Research Building, Morgantown, WV 26505, USA}
\author{Mohit Bhardwaj}
  \affiliation{Department of Physics, McGill University, 3600 rue University, Montr\'eal, QC H3A 2T8, Canada}
  \affiliation{McGill Space Institute, McGill University, 3550 rue University, Montr\'eal, QC H3A 2A7, Canada}
\author{Charanjot Brar}
  \affiliation{Department of Physics, McGill University, 3600 rue University, Montr\'eal, QC H3A 2T8, Canada}
  \affiliation{McGill Space Institute, McGill University, 3550 rue University, Montr\'eal, QC H3A 2A7, Canada}
\author{Tomas Cassanelli}
  \affiliation{Dunlap Institute for Astronomy \& Astrophysics, University of Toronto, 50 St.~George Street, Toronto, ON M5S 3H4, Canada}
  \affiliation{David A.~Dunlap Department of Astronomy \& Astrophysics, University of Toronto, 50 St.~George Street, Toronto, ON M5S 3H4, Canada}
\author{Pragya Chawla}
  \affiliation{Department of Physics, McGill University, 3600 rue University, Montr\'eal, QC H3A 2T8, Canada}
  \affiliation{McGill Space Institute, McGill University, 3550 rue University, Montr\'eal, QC H3A 2A7, Canada}
  \affiliation{Anton Pannekoek Institute for Astronomy, University of Amsterdam, Science Park 904, 1098 XH Amsterdam, The Netherlands}
\author{Fengqiu Adam Dong}
  \affiliation{Department of Physics and Astronomy, University of British Columbia, 6224 Agricultural Road, Vancouver, BC V6T 1Z1 Canada}
\author{Deborah Good}
  \affiliation{Department of Physics, University of Connecticut, 196 Auditorium Road, U-3046, Storrs, CT 06269-3046, USA}
  \affiliation{Center for Computational Astrophysics, Flatiron Institute, 162 5th Avenue, New York, NY 10010, USA}
\author{Victoria Kaspi}
  \affiliation{Department of Physics, McGill University, 3600 rue University, Montr\'eal, QC H3A 2T8, Canada}
  \affiliation{McGill Space Institute, McGill University, 3550 rue University, Montr\'eal, QC H3A 2A7, Canada}
\author{Adam E.~Lanman}
  \affiliation{Department of Physics, McGill University, 3600 rue University, Montr\'eal, QC H3A 2T8, Canada}
  \affiliation{McGill Space Institute, McGill University, 3550 rue University, Montr\'eal, QC H3A 2A7, Canada}
\author{Hsiu-Hsien Lin}
  \affiliation{Institute of Astronomy and Astrophysics, Academia Sinica, Astronomy-Mathematics Building, No. 1, Sec. 4, Roosevelt Road, Taipei 10617, Taiwan}
  \affiliation{Canadian Institute for Theoretical Astrophysics, 60 St.~George Street, Toronto, ON M5S 3H8, Canada}
\author{Bradley W.~Meyers}
  \affiliation{Department of Physics and Astronomy, University of British Columbia, 6224 Agricultural Road, Vancouver, BC V6T 1Z1 Canada}
\author{Aaron B.~Pearlman}
  \altaffiliation{McGill Space Institute Fellow}
  \altaffiliation{FRQNT Postdoctoral Fellow}
  \affiliation{Department of Physics, McGill University, 3600 rue University, Montr\'eal, QC H3A 2T8, Canada}
  \affiliation{McGill Space Institute, McGill University, 3550 rue University, Montr\'eal, QC H3A 2A7, Canada}
\author{Ue-Li Pen}
  \affiliation{Institute of Astronomy and Astrophysics, Academia Sinica, Astronomy-Mathematics Building, No. 1, Sec. 4, Roosevelt Road, Taipei 10617, Taiwan}
  \affiliation{Canadian Institute for Theoretical Astrophysics, 60 St.~George Street, Toronto, ON M5S 3H8, Canada}
  \affiliation{Canadian Institute for Advanced Research, 180 Dundas St West, Toronto, ON M5G 1Z8, Canada; }
  \affiliation{David A.~Dunlap Department of Astronomy \& Astrophysics, University of Toronto, 50 St.~George Street, Toronto, ON M5S 3H4, Canada}
  \affiliation{Perimeter Institute of Theoretical Physics, 31 Caroline Street North, Waterloo, ON N2L 2Y5, Canada}
\author{Emily Petroff}
  \affiliation{Department of Physics, McGill University, 3600 rue University, Montr\'eal, QC H3A 2T8, Canada}
  \affiliation{McGill Space Institute, McGill University, 3550 rue University, Montr\'eal, QC H3A 2A7, Canada}
  \affiliation{Anton Pannekoek Institute for Astronomy, University of Amsterdam, Science Park 904, 1098 XH Amsterdam, The Netherlands}
\author{Ziggy Pleunis}
  \affiliation{Dunlap Institute for Astronomy \& Astrophysics, University of Toronto, 50 St.~George Street, Toronto, ON M5S 3H4, Canada}
\author{Masoud Rafiei-Ravandi}
  \affiliation{Department of Physics, McGill University, 3600 rue University, Montr\'eal, QC H3A 2T8, Canada}
  \affiliation{McGill Space Institute, McGill University, 3550 rue University, Montr\'eal, QC H3A 2A7, Canada}
\author{Mubdi Rahman}
  \affiliation{Sidrat Research, PO Box 73527 RPO Wychwood, Toronto, ON M6C 4A7, Canada}
\author{Pranav Sanghavi}
  \affiliation{Lane Department of Computer Science and Electrical Engineering, 1220 Evansdale Drive, PO Box 6109, Morgantown, WV 26506, USA}
  \affiliation{Center for Gravitational Waves and Cosmology, West Virginia University, Chestnut Ridge Research Building, Morgantown, WV 26505, USA}
\author{Paul Scholz}
  \affiliation{Dunlap Institute for Astronomy \& Astrophysics, University of Toronto, 50 St.~George Street, Toronto, ON M5S 3H4, Canada}
\author{Kaitlyn Shin}
  \affiliation{MIT Kavli Institute for Astrophysics and Space Research, Massachusetts Institute of Technology, 77 Massachusetts Ave, Cambridge, MA 02139, USA}
  \affiliation{Department of Physics, Massachusetts Institute of Technology, 77 Massachusetts Ave, Cambridge, MA 02139, USA}
\author{Seth Siegel}
  \affiliation{Department of Physics, McGill University, 3600 rue University, Montr\'eal, QC H3A 2T8, Canada}
\author{Kendrick M.~Smith}
  \affiliation{Perimeter Institute of Theoretical Physics, 31 Caroline Street North, Waterloo, ON N2L 2Y5, Canada}
\author{Ingrid Stairs}
  \affiliation{Department of Physics and Astronomy, University of British Columbia, 6224 Agricultural Road, Vancouver, BC V6T 1Z1 Canada}
\author{Shriharsh P.~Tendulkar}
  \affiliation{Department of Astronomy and Astrophysics, Tata Institute of Fundamental Research, Mumbai, 400005, India}
  \affiliation{National Centre for Radio Astrophysics, Post Bag 3, Ganeshkhind, Pune, 411007, India}
\author{Keith Vanderlinde}
  \affiliation{Dunlap Institute for Astronomy \& Astrophysics, University of Toronto, 50 St.~George Street, Toronto, ON M5S 3H4, Canada}
  \affiliation{David A.~Dunlap Department of Astronomy \& Astrophysics, University of Toronto, 50 St.~George Street, Toronto, ON M5S 3H4, Canada}
\newcommand{\allacks}{
A.B.P. is a McGill Space Institute (MSI) Fellow and a Fonds de Recherche du Quebec -- Nature et Technologies (FRQNT) postdoctoral fellow.
C.L. was supported by the U.S. Department of Defense (DoD) through the National Defense Science \& Engineering Graduate Fellowship (NDSEG) Program
E.P. acknowledges funding from an NWO Veni Fellowship.
F.A.D is funded by the UBC Four Year Doctoral Fellowship.
FRB research at UBC is funded by an NSERC Discovery Grant and by the Canadian Institute for Advanced Research. The CHIME/FRB baseband system is funded in part by a Canada Foundation for Innovation JELF grant to IHS.
J.M.P is a Kavli Fellow.
K.M.B. is supported by an NSF grant (2006548, 2018490)
K.S. is supported by the NSF Graduate Research Fellowship Program.
K.W.M. is supported by NSF grants 2008031 and 2018490.
M.B. is supported by an FRQNT Doctoral Research Award.
M.D. is supported by a Killam Fellowship, CRC Chair, NSERC Discovery Grant, CIFAR, and by the FRQNT Centre de Recherche en Astrophysique du Qu\'ebec (CRAQ).
P.S. is a Dunlap Fellow and an NSERC Postdoctoral Fellow. 
V.M.K. holds the Lorne Trottier Chair in Astrophysics \& Cosmology, a Distinguished James McGill Professorship, and receives support from an NSERC Discovery grant (RGPIN 228738-13), from an R. Howard Webster Foundation Fellowship from CIFAR, and from the FRQNT CRAQ.
We acknowledge the support of the Natural Sciences and Engineering Research Council of Canada (NSERC), [funding reference number RGPIN-2019-067, CRD 523638-201, 555585-20] We receive support from Ontario Research Fund—research Excellence Program (ORF-RE), Canadian Institute for Advanced Research (CIFAR), Thoth Technology Inc, Alexander von Humboldt Foundation, and the Ministry of Science and Technology(MOST) of Taiwan(110-2112-M-001-071-MY3). 
Z.P. is a Dunlap Fellow.
}

\date{\today}

\begin{abstract}
	Fast radio bursts (FRBs) represent an exciting frontier in the study of gravitational lensing, due to their brightness, extragalactic nature, and the compact, coherent characteristics of their emission. 
	In a companion work~\citep{kader2022high}, we use a novel interferometric method to search for gravitationally lensed FRBs in the time domain using bursts detected by CHIME/FRB. There, we dechannelize and autocorrelate electric field data at a time resolution of \SI{1.25}{\nano\second}. This enables a search for FRBs whose emission is coherently deflected by gravitational lensing around a foreground compact object such as a primordial black hole (PBH). Here, we use our non-detection of lensed FRBs to place novel constraints on the PBH abundance outside the Local
    Group. We use a novel two-screen model to take into account decoherence from scattering screens in our constraints. Our constraints are subject to a single astrophysical model parameter -- the effective distance between an FRB source and the scattering screen, for which we adopt a fiducial distance of \SI{1}{\parsec}. We find that coherent FRB lensing is a sensitive probe of sub-solar mass compact objects. Having observed no lenses in \revCL{$172$ bursts from $114$ independent
    sightlines through the cosmic web},
    we constrain the fraction of dark matter made of compact objects, such as PBHs, to be $f \lesssim 0.8$, if their masses are $\sim 10^{-3} M_{\odot}$.
\end{abstract}

\maketitle


\section{Introduction}

Gravitational lensing occurs when spatially-inhomogeneous distributions of mass perturb spacetime and allow 
the light from background sources to take multiple paths on their way to \revCL{the observer}.
Since serving as one of the first historic confirmations of general relativity~\citep{dyson1920determination}, gravitational lensing has become firmly established as a powerful tool for astrophysics and cosmology. 
It has been used to measure the mass of galaxy clusters~\citep{zwicky1937nebulae,walsh1979twin,hoekstra2013masses}, to probe the substructure of dark matter halos~\citep{moore1999dark,mao1999evidence}, and as an independent probe of $H_0$ with time-delay cosmography~\citep{refsdal1964possibility,suyu2010dissecting}. 
Using large, time-domain surveys, the frontier in gravitational lensing has turned to searching for lensed transients such as supernovae~\citep{kelly2015multiple,goobar2017multiply,zumalacarregui2018limits}, gamma-ray bursts (GRBs)~\citep{nowak1994can, ji2018strong, paynter2021evidence, wambsganss1993method}, gravitational waves~\citep{nakamura1998gravitational,2018ngprecise}, and fast radio bursts (FRBs)~\citep{li2018strongly, munoz2016lensing, farah2018frb,day2020high,sammons2020first}. These searches can yield robust constraints on the
abundance of dark compact objects such as primordial black holes (PBHs)~\citep{mao2012astrophysical,munoz2016lensing,carr2020primordial,green2020primordial}.

PBHs may make up a large fraction of the dark matter, and could offer new observational handles on early-universe inflationary physics~\citep{green2020primordial,carr2020primordial}. However they \revCL{are} notoriously difficult to probe since their mass function is unknown and can span many orders of magnitude depending on their formation and evolution history. One recent constraint on sub-solar mass PBHs comes from the observation of an optical microlensing event towards
M31~\citep{niikura2019microlensing}. While these constraints are stringent, they apply only to PBHs along the line of sight towards M31. To constrain the cosmological abundance of PBHs, more distant backlights must be used. The lack of microlensed Type Ia supernovae in the local Universe implies that if the PBH dark matter has a mass function peaked at some central mass $10^{-2} M_{\odot} \lesssim M_{c} \lesssim 10^4 M_{\odot}$, the fraction of dark matter within compact lenses at low redshifts can be constrained to be $f < 0.35$~\citep{zumalacarregui2018limits}.

In this work, we analyze and interpret the results of a novel time domain search for lensed FRBs in a sample of bursts detected by the CHIME/FRB experiment~\citep{FRBSystemOverview,chimefrbcatalog1}. FRBs~\citep{lorimer2007bright,petroff2019fast,cordes2019fast} are millisecond-duration radio transients whose brightness,
compactness, and all-sky rate of $\sim 10^4$ Gpc$^{-3}$ yr$^{-1}$~\citep{lu2019implications,cordes2019fast,ravi2019prevalence,oguri2019strong} make them
outstanding backlights for time-domain lensing science.

Time-domain lensing searches typically look for multi-peak light curves which arise from different lensing time delays and different magnification ratios between images. Perhaps the most difficult question in any time-domain search for lensed transients~\citep{paczynski1986gamma,paczynski1987gravitational,ougolnikov2001search,hurley2019search,paynter2021evidence} is: How can temporal structure in transient light curves induced by gravitational lensing be conclusively distinguished
from intrinsically complex temporal structures? In past searches, detailed statistical analysis of transient morphology in multiple observing bands is often used to
answer this question.

In our search we apply a coherent correlation algorithm, detailed in a companion work~\citep{kader2022high}, which breaks the degeneracy between pulse morphology and gravitational lensing. This uses the fact that gravitational lensing coherently applies a delay between the two images, a measurable effect in the wave field domain. Our coherent correlation algorithm is similar to that used in very long baseline interferometry, which relies on the presence of phase-preserving records of electric field data (hereafter referred to as ``baseband data''). We refer to this technique as {\it FRB gravitational-lens interferometry}. In our search we dechannelize CHIME/FRB electric field data and autocorrelate it to search for coherently-delayed copies of the same signal. Using electric field rather than intensity information improves time-lag resolution from the variability timescale of the transient (milliseconds) to the Nyquist limit of the telescope (nanoseconds).
 \revCL{This improves our ability to probe low-mass objects by orders of magnitude.}
Our coherent search method improves sensitivity to fainter images compared to incoherent methods, and gives access to shorter delay timescales (lower mass scales). This complements the incoherent method of previous works based on the intensity light curves~\citep{munoz2016lensing,sammons2020first}. This allows us to constrain lensing delays of $10^{-9}-10^{-1}$ sec, corresponding to PBHs in the mass range of $10^{-4}-10^4 M_{\odot}$. These coherent techniques open up the exciting possibility for
high time-resolution studies of FRB sources, observation of wave-optical effects in gravitational lensing~\citep{nakamura1998gravitational,jow2020wave,katz2020looking}, and even so-called ``real-time'' cosmology~\citep{eichler2017nanolensed,zitrin2018observing,pearson2020searching,wucknitz2021cosmology}.

\section{Search Description}
The sensitivity of any microlensing search can be characterized by calculating the expected number of lensing events for given survey parameters. The observed number of lensing events, $k$, is connected to the theoretical lensing rate $\lambda$, through Poisson statistics. The traditional formalism for calculating $\lambda$ (also known as the lensing optical depth) was developed for optical microlensing
surveys~\citep{paczynski1986gravitational,paczynski1987gravitational,griest1991galactic,griest1991gravitational}. We briefly review the traditional formalism and show how we extend it to handle a detailed description of the sensitivity of our time-domain search.

Traditionally, three numbers quantify the sensitivity of a survey to gravitational lensing: the minimum and maximum delay timescales
$\tau_{\rm min},\tau_{\rm max}$ for which the survey is sensitive, as well as the \revCL{minimum} detectable flux magnification ratio between the two images. Early work on FRB lensing adapted this formalism for parameterizing time-domain surveys~\citep{munoz2016lensing,laha2018lensing, sammons2020first}. $\tau_{\rm min}$ is typically set to the variability timescale of the transient, and $\tau_{\rm max}$ is taken to be the maximum lensing delay detectable in a given search (often the duration of data capture). 

For a lens to be detectable by a given search, the lensing delay must fall between $\tau_{\rm min}$ and $\tau_{\rm max}$, and the double image must be sufficiently bright to be detected. The latter criterion is typically written as a constraint on what flux ratios are detectable. If the flux ratio (often denoted $\varepsilon$) is taken by convention to be greater than 1 as in~\citep{munoz2016lensing,laha2018lensing}, the flux criterion is written as $1 < \varepsilon < \varepsilon_{\rm max}$ for some specified choice of $\varepsilon_{\rm max}$. For
example, some works~\citep{munoz2016lensing,laha2018lensing} assume that lensing events are detectable when the dimmer image is no more than 5 times dimmer than the main image, requiring that $1 < \varepsilon < \varepsilon_{\rm max} = 5$. 
A more realistic criteria is that $1 < \varepsilon < \varepsilon_{\rm max} = 1/3 \times \mathrm{S/N}$, where S/N refers to the signal-to-noise ratio at which the burst can be detected in autocorrelation~\citep{sammons2020first}. This captures the fact that for a brighter burst, images with fainter flux ratios may be detected. 

In our coherent search~\citep{kader2022high}, we depart from this convention in two ways. \revCL{First, we refer to the ratio of the \textit{wavefield amplitudes} between the two images as $\varepsilon$, and the corresponding flux ratio as $\varepsilon^2$. Second, we take the reciprocal convention where the smaller of the two amplitudes is in the numerator. All of our flux ratios are therefore less than 1. In this convention, the search sensitivity can be expressed in the form
$\varepsilon_{\mathrm min}^2 < \varepsilon^2 < 1$.} Our search parameterizes the search sensitivity in a more detailed way than previous works. Instead of assuming a constant threshold $\varepsilon^2_{\mathrm{min}}$, we allow the \revCL{threshold} to vary as a function of delay: $\varepsilon_{\rm min}^2(\tau) < \varepsilon^2 < 1$, where $\tau$ is the trial delay over which we search for lenses. The motivation for this parameterization comes from our novel search algorithm, which measures $\varepsilon^2_{\rm min}(\tau)$.

\subsection{Search Data Products}
We briefly review the details of our search algorithm here but refer the reader to~\citet{kader2022high} for an in-depth discussion. The input data to the search is channelized baseband data, produced by forming a voltage beam towards the best-fit sky position of the source~\citep{michilli2020analysis}. We measure the time-lag autocorrelation function (ACF) of the FRB by combining the baseband data from 1024 frequency channels, each with a time resolution of \SI{2.56}{\micro\second}, into a single voltage time stream $V_P(t)$ with a time resolution of \SI{1.25}{\nano\second} independently for the two polarizations of our telescope ($P = X$,$Y$). $V_X(t)$ and $V_Y(t)$ represent the electric field projected onto the two telescope polarizations. Then, each timestream is windowed by multiplying it by a function $W_P^2(t)
\propto \mathrm{S/N_P}(t)$, where $\mathrm{S/N_P}[t]$ is proportional to the normalized flux of the burst
as detected in each telescope polarization over time~\citep{leung2020synoptic}. We search for echoes by time-lag correlating the product $W_P^2(t)V_P(t)$ against $V_P(t)$, shifted by many ($\sim 10^8$) trial delays $\tau$. This yields two ACFs: $C_X(\tau)$ and $C_Y(\tau)$, which are converted into measurements of $\varepsilon_X(\tau)$ and $\varepsilon_Y(\tau)$ (see Appendix C of~\citep{kader2022high}). In $\varepsilon$ units, we measure $\varepsilon^2_{\mathrm{min}}(\tau)$ as follows. We define $\mu_P(\tau)$ as the
mean of $\varepsilon(\tau)$, measured when the pulse is off. $\mu_P(\tau)$ has the property that it differs from zero for $|\tau| \lesssim \SI{300}{\nano\second}$ due to instrumental reflections, seen even when the pulse is off. We therefore subtract this background for short lags. Then, we have 
\begin{equation}
    \varepsilon_{\rm min}^2(\tau) = (\varepsilon_X(\tau) + \varepsilon_Y(\tau))^2 < 1.
    \label{eq:acf_to_eps}
\end{equation}
We allow $\varepsilon^2$ to vary as a function of $\tau$ for the following reasons. First, different systematics are present in the ACF of at different time delays. For example, $\sim$kilohertz bandwidth radio-frequency interference (RFI) present in our data is often relevant at millisecond delay scales, and less so at shorter delays. Second, we experience a sensitivity drop for long time-lags due to the look-elsewhere effect (see discussion in Ref.~\citep{kader2022high}, Section IV). Third, in the presence of a fixed amount of pulse broadening arising from a scattering screen, we expect interferometric lensing to decohere more easily for larger lens masses (and therefore larger time-lags) than at smaller lenses and smaller lags (see Sec~\ref{sec:screens}).

\section{Lensing Event Rate}

To convert our search sensitivity into constraints on primordial black holes, it is first necessary to define the conditions under which the alignment of a lens with an FRB creates a detectable double image. This requires assuming a lens model which specifies the mass distribution within the lens plane. \revCL{Each distinct FRB source can then be thought of as an independent ``sightline'': a pencil beam whose endpoints are at Earth and the FRB source, and whose beam size is set by the Einstein
radius.} For each sightline (indexed hereafter by $j$), we integrate over different lensing geometries (Sec.~\ref{sec:geo}) to get the total lensing cross-section $\sigma_{ij}$, where $i$ indexes possible ranges of time-lags. Next, we must integrate over possible lens redshifts $z_\mathrm{L}$. Doing this requires translating redshifts to cosmological distances through $H(z) = H_0\sqrt{\Omega_m (1+z)^3 + \Omega_{\Lambda}}$ and the source redshift $z_{S,j}$, which must be inferred for unlocalized FRBs (Sec.~\ref{sec:z_lens}) from their DM. Finally, we must integrate over the lens mass
function $dn_c/dM$ (Sec.~\ref{sec:mass_function}), which depends on $f$.
The method is summarized by Eq.~\ref{eq:n_delta}, which represents the expected number of lensing events at time lags within some $\tau_{i}$ and $\tau_{ i+1}$ for a sightline $j$. We call this $\lambda_{ij}$ (for the $i$-th time-lag bin and the $j$th FRB source).
\begin{equation}
	\lambda_{ij} \equiv \int_0^{\infty}~dM \dfrac{dn_c}{dM} \int_0^{z_{S,j}} dz_\mathrm{L} \dfrac{c(1 + z_\mathrm{L})^2}{H(z_\mathrm{L})}\sigma_{ij}.
\label{eq:n_delta}
\end{equation} 
We sum over $i$ and $j$ to calculate the total event rate (Sec.~\ref{sec:combining}); this allows us to set upper limits on $f$, the fraction of dark matter that is made of compact lenses such as PBHs.

\subsection{Possible Lensing Geometries}
\label{sec:geo}
\begin{figure}
    \includegraphics{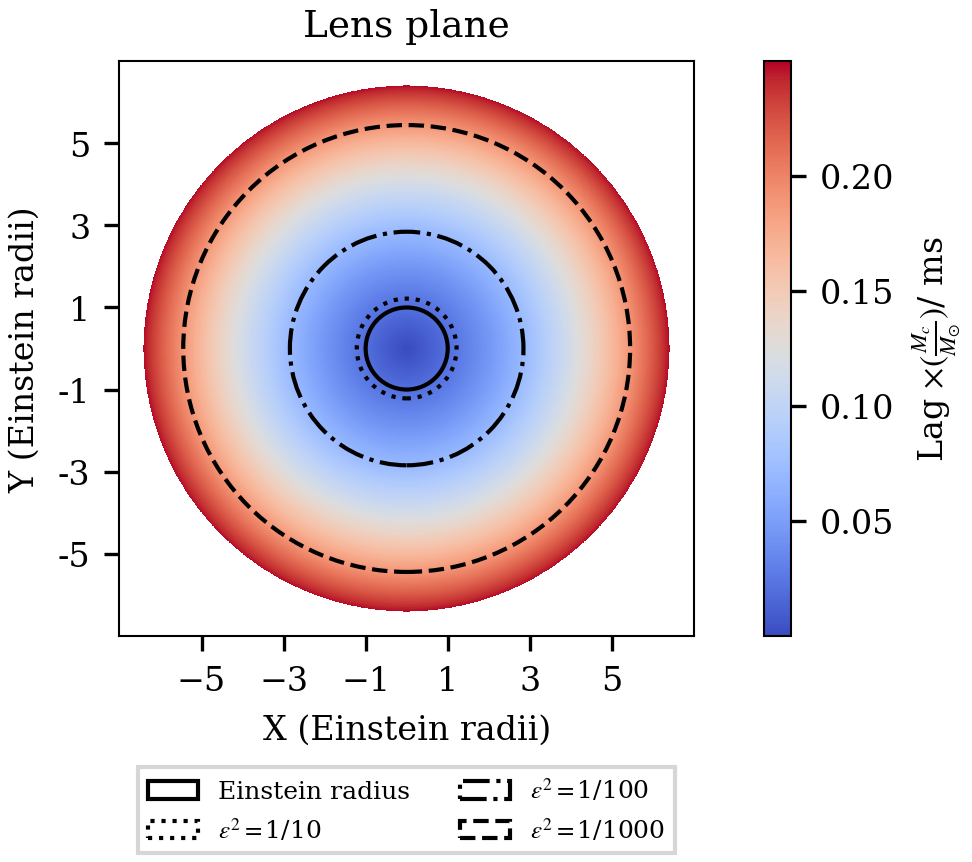}
    \caption{Our schematic depiction of the lens plane, with coordinates centered on the source's unlensed position, and transverse distances measured in units of Einstein radii. We shade the delay between images as a function of lens's transverse position in the lens plane (colored disk). A time-delay based detection search space constrains possible lens positions to an annulus within the plane. A flux-based detection threshold, parameterized by $\varepsilon^2_{\mathrm{min}}$, further
	constrains the annulus's outer boundary via Eq.~\ref{eq:ymax_flux} (dotted boundaries). The lensing cross-section $\sigma$ can be understood as the area of the annulus that satisfies both the flux- and time-delay based detection thresholds (Eq.~\ref{eq:donut}).}
    \label{fig:donut}
\end{figure}

In our search, we aim to detect multiple temporally-resolved FRB images from a compact object acting as a lens. For compact objects, it is a good approximation to use a point-mass lens model, which predicts two images except in the edge case of an Einstein ring. We wish to integrate over all possible geometries (parametrized by the different possible impact parameters $b$) which could produce a lensing event. To do so, we briefly summarize the relationship between the astrophysical parameters (the lens mass, redshift, and impact parameter) and the observables (the Shapiro delay and the flux magnification ratio $\varepsilon^2$) for the point lens model. We introduce the dimensionless impact parameter $y = b/R_E$, where $b$ is the physical impact
parameter of the source in the lens plane at the lens redshift $z_\mathrm{L}$, and where $R_E(M,z_\mathrm{L},z_S)$ is the Einstein radius of a lens with some mass $M$ at redshift $z_\mathrm{L}$, magnifying an FRB at $z_{S}$ (see also Fig.~\ref{fig:two_screen}).
With these definitions the differential Shapiro delay between images is given by (see also~\citep{munoz2016lensing})
\begin{equation}
    \tau = \dfrac{2 R_s(1 + z_\mathrm{L})}{c} g(y)
    \label{eq:shapiro_delay}
\end{equation}
where $R_s$ is the Schwarzschild radius of a putative lens of mass $M$, $z_\mathrm{L}$ is the lens redshift, and
\begin{equation}
    g(y) = \dfrac{1}{2}(y\sqrt{y^2 + 4}) + \log\left(\dfrac{\sqrt{y^2 + 4} + y}{\sqrt{y^2 + 4} - y}\right).
\end{equation}
We visualize Eq.~\ref{eq:shapiro_delay} as a function of the lens position in the lens plane in Fig.~\ref{fig:donut}.
For a given impact parameter $y$, the magnification ratio between the two images is
\begin{equation}
    y(\varepsilon^2) = \sqrt{\sqrt{\varepsilon^2} + 1/\sqrt{\varepsilon^2} - 2}.
    \label{eq:ymax_flux}
\end{equation}
To detect a lens, we require that the dimmer image be bright enough to be detected in voltage cross-correlation, and that the Shapiro delay between images falls within the valid delay range over which we are sensitive. We can express these criteria as upper and lower bounds on $y$, respectively.
The minimum impact parameter $y_{\rm min}$ at which a lens is detectable is determined by the minimum resolvable Shapiro delay between images (i.e. plugging in $\tau_{\rm min}$ in Eq.~\ref{eq:shapiro_delay}). The maximum impact parameter is a function of $\varepsilon^2$ only. Several flux thresholds are drawn in our depiction of the lens plane Fig.~\ref{fig:donut}.
However, in our search, $\varepsilon^2$ varies as a function of lag. We partition the full range of accessible delays (in our case, $10^{-9}-10^{-1}$ seconds) into logarithmically-spaced bins, indexed by $i$, with boundaries $\tau_{\mathrm i} < \tau < \tau_{\mathrm{i}+1}$. Within each logarithmically-spaced lag range $[\tau_{\rm i},\tau_{\rm i+1})$ with corresponding values of $\varepsilon^2(\tau)$, we take 
\begin{equation} 
y_{\rm min,i} = g^{-1}\left( \dfrac{c\tau_{\rm i}}{2 R_s (1 + z_\mathrm{L})} \right)
    \label{eq:ymin}
\end{equation}
and 
\begin{equation} 
y_{\rm max,ij} = \min_{\tau \in (\tau_{\rm i},\tau_{\rm i+1})} \left\{
g^{-1}\left( \dfrac{c\tau_{\rm i+1}}{2 R_s (1 + z_\mathrm{L})} \right), y(\varepsilon^2_j(\tau)) \right\} .
\label{eq:ymax}
\end{equation}

In Fig.~\ref{fig:donut} we plot the contribution of different delay timescales to the lensing cross-section, which may be geometrically interpreted as an annulus in the lens plane with boundaries $y_{\mathrm{min,i}}$ and $y_{\mathrm{max,ij}}$ due to detectability (Eqs.~\ref{eq:ymax_flux},\ref{eq:ymin},\ref{eq:ymax}). We see that smaller misalignments from $b=0$ (the center of the annulus) correspond to shorter Shapiro delays and larger misalignments ($b >> R_E$) produce long delays and extreme flux ratios. The area of the annulus is the cross-section to lensing: 
\begin{equation} 
	\sigma_{ij} = \pi R_{E,j}^2 (y_{\mathrm{max,i}}^2 - y_{\mathrm{min,ij}}^2).\label{eq:donut}
\end{equation} 

\subsection{Distance Inference}\label{sec:distances}
\label{sec:z_lens}
It is difficult to infer the distance of FRBs without independent redshift measurements. One promising proxy for distance is the amount of dispersion in the dynamic spectrum of the FRB, accumulated as the FRB passes through the intervening cold plasma of the intergalactic medium (IGM) on its way towards the observer. The dispersion is quantified by the dispersion measure (DM) and is readily measurable, making this an attractive approach. 

However, inferring distances from DMs has many biases and
uncertainties, which depend on factors such as the FRB luminosity function, the survey depth/field of view, the spectral index of the FRB emission, and contributions to the DM which do not correlate with distance (e.g. local environments or intervening haloes). When these effects are taken into account, it has been shown that under certain circumstances, large DMs are a poor proxy for distance -- i.e., that the highest DM events in an FRB survey may not be the most distant~\citep{james2021z}. To minimize these biases, bursts in our sample are not selected on the basis of properties like their DM or brightness. Instead, we select bursts only on the criteria that baseband data were collected and processed. For baseband data to be collected, the FRB must have a minimum S/N of $\approx 15$ to reduce the volume of false positives collected; this threshold has varied between 12 and 20 over the course of CHIME/FRB's operation. In addition, detected FRBs have a maximum DM of $\approx \SI{1000}{\parsec~\centi\meter^{-3}}$, which is imposed by the memory size of the ring buffer within the CHIME
correlator). This is a factor of 2-3 below the highest-DM events observed by the CHIME/FRB instrument for which we expect high-DM selection bias to dominate.
The total amount of smearing is quantified by $\mathrm{DM}_{\mathrm{obs}}$ and can be written as a sum of contributions from the Milky Way ($\mathrm{DM}_{\mathrm{MW}}$), the intergalactic medium ($\mathrm{DM}_{\rm IGM}(z_S)$), and the host galaxy ($\mathrm{DM_{host}}$). as shown in Eq.~\ref{eq:dm_eg}. We determine $z_S$ by first solving Eq.~\ref{eq:dm_eg} for $\mathrm{DM}_{\rm IGM}$, taking $\mathrm{DM_{mw}}$ to be the NE2001 expectation along the line of sight. 
\begin{equation}
\label{eq:dm_eg}
    {\rm DM}_{\rm obs} = {\rm DM}_{\rm mw} + {\rm DM}_{\rm IGM}(z_S) +  {\rm DM}_{\rm host}(z_S)
\end{equation}
We conservatively model $\mathrm{DM_{host}}$ as,
\begin{equation}
\label{eq:dm_host}
    {\rm DM}_{\rm host}(z_S) = \dfrac{117~\mathrm{pc~cm}^{-3}}{1 + z_S} .
\end{equation}
There are several reported values for the average $\mathrm{DM_{host}}$ in the literature; however, $117~\mathrm{pc~cm}^{-3}$ is the median value favored by an analysis of the luminosity function of CHIME-detected FRBs using CHIME/FRB Catalog 1, after correcting for known selection effects~\citep{shin2022luminosity}. This is consistent with the value reported for ASKAP
FRBs with a similar analysis ($145^{+64}_{-60} \textrm{pc~cm}^{-3}$)~\citep{james2021z}. In principle, drawing from a distribution of $\mathrm{DM_{host}}$ values around the median CHIME/FRB value would be most realistic. However, since intrinsic correlations between $\mathrm{DM_{host}}$ and other properties (e.g. the FRB's distance and brightness) are poorly constrained, we adopt the median value for all FRBs, and quantify uncertainties related to distance determination by exploring two astrophysically-motivated scenarios (see Sec.~\ref{sec:discussion}).

Once $\mathrm{DM}_{\rm host}$ is assumed for each FRB, we can infer $\mathrm{DM}_{\rm IGM}$. We invert the Macquart relation~\citep{macquart2020census} (Eq.~\ref{eq:macquart_relation}) to determine the source redshift $z_\mathrm{S}$. In Eq.~\ref{eq:macquart_relation}, we have approximated the Universe's chemical composition as $75\%$ hydrogen and $25\%$ helium by mass, both completely ionized. This leads to $n_{e,0} = 0.875 \Omega_b \rho_{crit}/m_p$. Throughout this work, we assume a Planck 2018~\citep{aghanim2018planck} cosmology with $H_0 = \SI{67.7}{\kilo\meter\per\second\per\mega\parsec}$ and $(\Omega_m,\Omega_b,\Omega_\Lambda) = (0.30966, 0.04897, 0.68884)$. 

\begin{equation}
    \label{eq:macquart_relation}
    {\rm DM_{IGM}} = \int_0^{z_S} \dfrac{c n_{e,0} (1+z)}{H_0 \sqrt{\Omega_m(1+z)^3 + \Omega_{\Lambda}}}~dz
\end{equation}

\subsection{Possible Lens Masses} 
\label{sec:mass_function}
The final integral in Eq.~\ref{eq:n_delta} is a marginalization over the unknown lens mass function $dn_c / dM$ which has units of comoving number density (denoted $n_c$) per unit mass. To constrain the abundance of compact lenses, we must assume a mass function of objects which produce the lensing events~\citep{green2016microlensing,carr2017primordial}. For PBHs,~\citet{green2016microlensing} suggests modeling the PBH function as a log-normal distribution peaked at some value of $\log_{10}(M_c/M_{\odot})$ and with some logarithmic width $\sigma$ measured in decades. 
For simplicity, we first consider the family of monochromatic mass functions (Eq.~\ref{eq:mass_function}):

\begin{equation}
    \dfrac{dn_c}{dM} = \dfrac{\rho_{\rm crit}}{M_c} \Omega_{\rm c} f(M_c) \delta(M - M_c).
    \label{eq:mass_function}
\end{equation}

This family of functions is parameterized solely by their central mass $M_c$, and have the property that if $f(M_c) = 1$, the total mass density is normalized to the cosmological dark matter density, i.e., 
\begin{equation}
    \int \dfrac{dn_c}{dM} M~dM = \rho_{\rm crit}\Omega_{\rm c}.
\end{equation}
The cosmological dark matter density $\Omega_{\rm c}$ is fixed at $\Omega_{\rm c} = \Omega_{m} - \Omega_{b} = 0.26069$~\citep{aghanim2018planck}. In practice, different formation scenarios give rise to both quasi-monochromatic $(\sigma / \log(M_c/M_{\odot}) \sim 1)$ and broad
($\sigma / \log_{10}(M_c/M_{\odot}) \gg 1$) mass functions~\citep{carr2017primordial,carr2020primordial}. However, since Eq.~\ref{eq:n_delta} is linear in $\dfrac{dn_c}{dM}$, and since an extended mass function is a linear superposition of delta functions, it is straightforward to translate our calculation for delta functions to extended PBH mass functions. This is necessary because extended PBH mass functions allow certain inflationary scenarios to evade current PBH constraints~\citep{cleese2015massive,green2016microlensing}.

\subsection{Combining Bursts}
\label{sec:combining}
After calculating the optical depth, it is necessary to combine many sightlines due to the rarity of lensing events. Only a handful of lensed supernovae have been conclusively detected~\citep{kelly2015multiple,goobar2017multiply}; detailed estimates suggest that lensed FRBs are similarly rare~\citep{oguri2019strong}. The occurrence of lensing events in lag range $i$ in the direction of any single sightline \revCL{$j$} can be thought of as a Poisson process with a low rate $\lambda_{ij}(M_c) \ll 1$. Since independent Poisson processes are additive, we define several event rates: the rate summed over lag bins but not sightlines ($\lambda_j$), the rate summed over sightlines but not lag bins ($\lambda_i$), and the total event rate for the entire search summed over sightlines and lag bins: 
\begin{equation}
    \lambda = \sum_{j} \lambda_{j} = \sum_{i} \lambda_{i} = \sum_{i} \sum_{j} \lambda_{ij}.
\label{eq:summed_sightlines}
\end{equation}

\begin{figure}
    \includegraphics{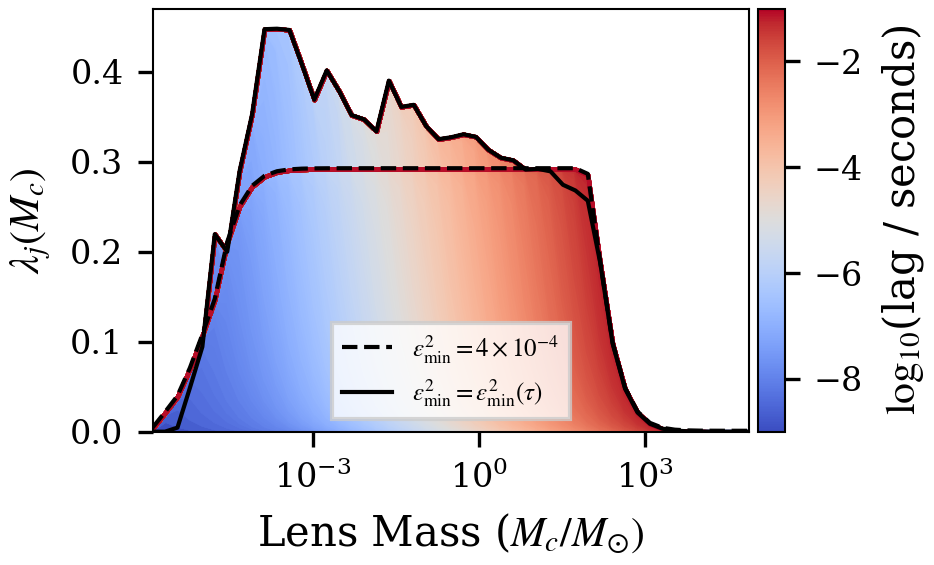}
    \caption{The expected lensing rate as a function of the lens mass for the sightline toward FRB 20191219F. The height of the curve can be interpreted as the Poisson rate of lensing events (i.e. the probability that the FRB is lensed) assuming that all dark matter is made up of compact
    lenses with mass $M_c$. For example, the probability of seeing a statistically-significant lensing signal if all the dark matter is composed of $\sim 10^{-1} M_{\odot}$ black holes is $\approx 0.6$. We calculate this rate via two methods, shown by the solid and dotted curves. Solid curve: sensitivity given by the ACF $\varepsilon^2(\tau)$ measured by our correlation algorithm. Dashed curve: sensitivity given by a constant fiducial value of $\varepsilon^2 = 10^{-4}$,
    shown to illustrate the difference with the approach taken by earlier work such as~\citep{sammons2020first}. Color shading denotes the additive contributions to the total probability from different time-delay scales. Relative to the constant-$\varepsilon^2$ case, the reduced event rate at short lags/low lens masses is because instrumental systematics in the delay spectrum at short delay scales ($\approx \SI{100}{\ns}$) degrade sensitivity. A similar reduction happens at long lags because of the large trials factor at large delay values (see text).}
    \label{fig:constant_rf}
\end{figure}

\begin{figure*}
    \includegraphics[width = 0.7\textwidth]{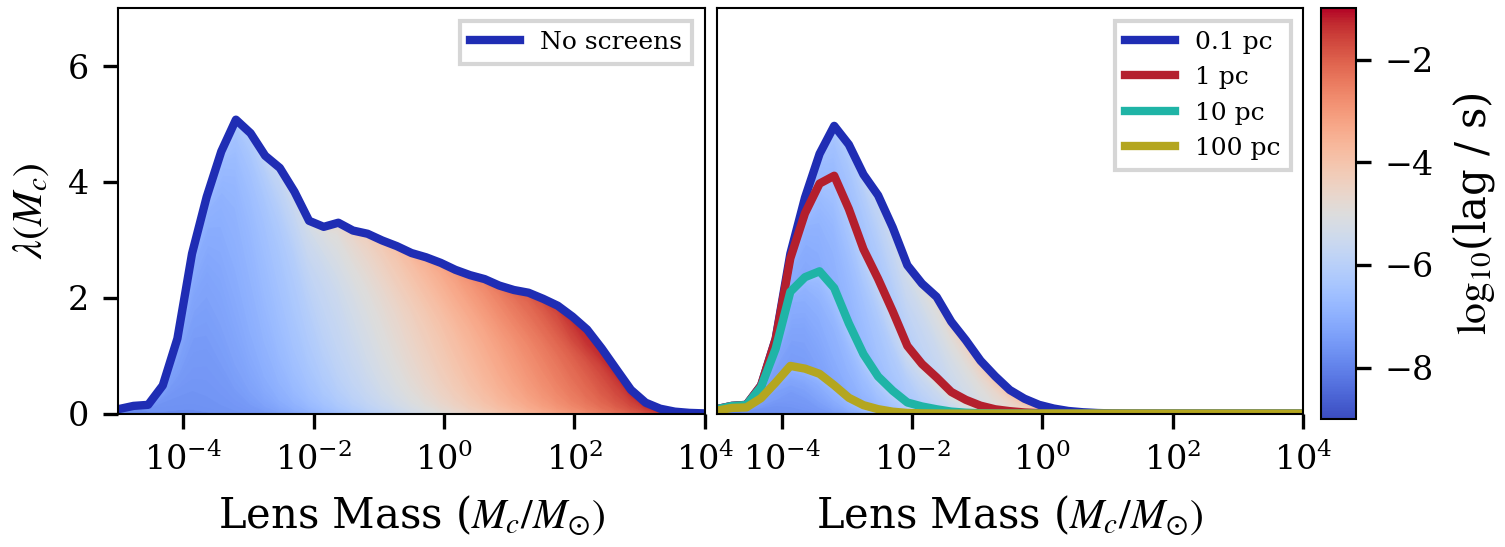}
    \caption{The expected lensing event rate for our full sample \revCL{of 114 FRB events assuming that all dark matter is composed of PBHs with mass $M_c$} (i.e., that $f  = 1$). Left: In the absence of plasma scattering screens which cause decoherence, the predicted lensing event rate extends over a wide range of PBH masses. Right: In the presence of plasma screens, the level of decoherence is sensitive to the screen's effective distance from the FRB source (different traces). This shows the impact of plasma scattering on coherent FRB lensing constraints.}
    \label{fig:optical_depth}
\end{figure*}

In Fig.~\ref{fig:constant_rf}, we visualize $\lambda_{ij}(M_c)$ (color shaded region) and its sum $\lambda_j(M_c)$ (thick line) for the sightline towards FRB 20191219F~\citep{leung2020synoptic}. We also compare our lag-dependent $\varepsilon_{\rm min}^2$ approach to the traditional approach (using an arbitrarily-chosen constant value of $\varepsilon_{\rm min}^2 = 10^{-4}$) in Fig.~\ref{fig:constant_rf}. Fig.~\ref{fig:optical_depth} shows the analogous quantities for the entire search
summed over all sightlines $i$. We visualize $\lambda_i(M_c)$ (shaded region) and its sum
$\lambda(M_c)$ (thick line) in Fig.~\ref{fig:optical_depth}. In both Fig.~\ref{fig:constant_rf} and Fig.~\ref{fig:optical_depth}, the color shading denotes the differential contributions of different delay timescales to their respective total rates. 

A complication arises from repeating FRBs. Repeat bursts from the same FRB source do not necessarily probe independent volumes of space. For low-mass PBHs with small Einstein radii, the motion of an FRB source (with respect to the Earth's rest frame) may move the sightline by many Einstein radii between successive bursts. If the lens and FRB do not reside in the same galaxy halo, their relative transverse velocity $v_{\perp}$ can be estimated by the velocity dispersion of a
typical galaxy cluster $\sigma\sim\SI{1000}{\km/\second}$~\citep{girardi1993velocity}. If we estimate the lensing distance to be $D_L D_{LS} / D_{S} \sim \SI{1}{\giga\parsec}$, the time it takes for the lens to move by one Einstein radius is
\begin{equation}
	T_{E} = \dfrac{R_E}{v_{\perp}} \sim \dfrac{R_\mathrm{s} D_L D_{LS}}{D_S \sigma} = \SI{14}{\year} \left( \dfrac{M}{M_{\odot}} \right)^{1/2}.
\end{equation}
It is evident that for some masses in the range $10^{-4}-10^{4} M_{\odot}$, the crossing time can be much shorter than the duration between two successive bursts from the same source (on the order of weeks or months). In this case, two repeat bursts illuminate disjoint sightlines through the cosmic web. However, in the opposite limit, the transverse motion of the FRB moves it by only a small fraction of its Einstein radius between successive bursts. In this case, repeat bursts from the same source illuminate the same
microlensing tube and cannot be counted independently in the total optical depth to lensing. To be conservative, we take only the brightest burst from each repeating FRB source for the best measurement of $\varepsilon^2(\tau)$ along that sightline. Combining information from repeat bursts is in principle possible by e.g., stacking the measured ACF over many bursts~\citep{krochek2021constraining}. However, at the nanosecond time resolution of our search, changes in the lensing delay over time (a so-called ``delay-rate'') must be taken into account to not wash out the signal from the stacking procedure. Hence, we defer an optimal treatment incorporating stacking to future work.

\section{Fundamental Limitations}
While our interferometric method allows us to break the degeneracy between pulse morphology and gravitational lensing, the lensing signal -- an interferometric fringe -- is also fragile and demands careful consideration of all possible sources of decoherence which could explain a non-detection. Wave optics effects, finite source size, and scattering screens all may result in a non-detection even when a lens is present. 

When the Schwarzschild radius of the lens is smaller than the wavelength of light, the light propagation is unaffected\citep{gould1992femtolensing,nakamura1998gravitational,nakamura1999wave,jow2020wave}. This imposes a low-mass sensitivity cutoff. For our observing frequencies we are only sensitive when $M > 1.5 (1 + z_\mathrm{L}) \times 10^{-4} M_{\odot}$~\citep{katz2020looking}. To overcome this, searches for FRB lensing should be conducted at higher observing frequencies \revCL{in order to probe very low-mass objects}.

A second important consideration for our search is the effect of finite source size~\citep{jow2020wave}. The astrophysics here are similar in spirit to femtolensing constraints from GRBs~\citep{nemiroff2001limits,gould1992femtolensing}, which were thought to apply to black holes of $M < 10^{-13} M_{\odot}$. However, GRB emission from an extended source is angularly incoherent when averaged over the size of the source, washing out the correlation
signal for all but the smallest GRBs emission regions~\citep{katz2018femtolensing}. The physics of this effect is similar to very long baseline interferometry experiments which ``resolve out'' extended sources on sufficiently long baselines, suppressing the cross-correlation fringes. This  invalidates PBH constraints from coherent femtolensing and one might wonder whether a similar concern applies to coherent FRB lensing constraints. For our coherent lag-correlation pipeline~\citep{kader2022high} to
detect a lensing event, the FRB emission region must appear as a point source as viewed with the resolving power of the gravitational lens.

An intuitive estimate (ignoring redshift effects) goes as follows. If the lens receives light of frequency $\nu_{\rm obs}$, its wavelength is $c / \nu_{\rm obs}$. If the lens is of size $R_{\mathrm{lens}}$, the FRB emission does not appear pointlike (i.e. is resolved) if $c / (\nu_{\rm obs} R_{\mathrm{lens}}) \sim r_{em} / D_{LS}$ where $r_{em}$ is the size of the emission region, and $D_{LS}$ is the distance from the lens to the source. For a point mass gravitational lens, $R_{\mathrm{lens}}$ can be approximated as the Einstein radius $R_E = \sqrt{2 R_s D_L D_{LS} / D_S}$ where $R_s$ is the Schwarzschild radius and $D_L,D_S$ are the angular diameter distance to the lens and the source respectively. 
Taking $D_{LS}, D_L,$ and $D_S$ to be on the order of 1 Gpc and $r_{em} = \SI{3e8}{\meter}$ (see below for justification of this choice), we conclude that a massive lens is needed to resolve the source: $M \sim 3\times 10^4 M_{\odot}$. This places a high-mass cutoff on our lensing constraints. For smaller masses the emission region will remain angularly coherent, unresolved by the gravitational lens. The maximum mass accessible scales as $\nu_{obs}^{-1}$. From this consideration, observing at lower frequencies is advantageous.

The size of the emission region can be estimated from the variability timescale of the transient, and fortunately, FRBs have much shorter variability timescales~\citep{cho2020spectropolarimetric,day2020high} than GRBs. 
The association of an FRB-like radio burst with the Galactic magnetar SGR 1935+2154~\citep{SGR,bochenek2020fast} is very strong evidence for a compact origin for at least some extragalactic FRBs. 
Mechanisms by which FRB emission is produced can be grouped into two broad categories: ``close-in'' models in which the bursts are produced in the magnetosphere (at radii of hundreds of kilometers) and shock models where the bursts are produced ``far-out'' from the central engine~\citep{metzger2019fast,plotnikov2019synchrotron}. The observation of diverse polarization properties including
millisecond-variability in the polarization angle~\citep{luo2020diverse}, long-term evolution, and significant circular polarization~\citep{xu2021fast} 
observed in a handful of FRBs tentatively challenge the latter class of models, though it is still a matter of intense debate.

Theories where the burst is emitted from the magnetosphere~\citep{kumar2017fast,zhang2017cosmic,yang2018bunching} involve distances of hundreds of neutron star radii and are significantly more compact than the synchrotron maser shock models. Shock models involve Lorentz factors of $\Gamma \sim 10^2$~\citep{waxman2017origin,plotnikov2019synchrotron,metzger2019fast}. For a millisecond-duration FRB, this corresponds to emission region size of $\lesssim 2\times10^{8}\Gamma$ cm~\citep{khangulyan2022fast}. 

In any of these FRB emission scenarios, the apparent angular size of the FRB emission region is quite compact. In the scenario where the FRB emission looks like a point source as resolved by a putative gravitational lens, the rate shown in the left panel of Fig.~\ref{fig:optical_depth} applies.
However, observations of pulsars and FRBs routinely indicate the presence of interstellar scattering along the line of sight. Interstellar scattering increases the effective angular size of apparent point sources at radio wavelengths, akin to atmospheric seeing for optical observations. This will be the topic of the next section, where we consider the effects of angular broadening due to scattering on our results.
\label{sec:source_size}

\begin{figure*}
    \includegraphics[width = 0.9\textwidth]{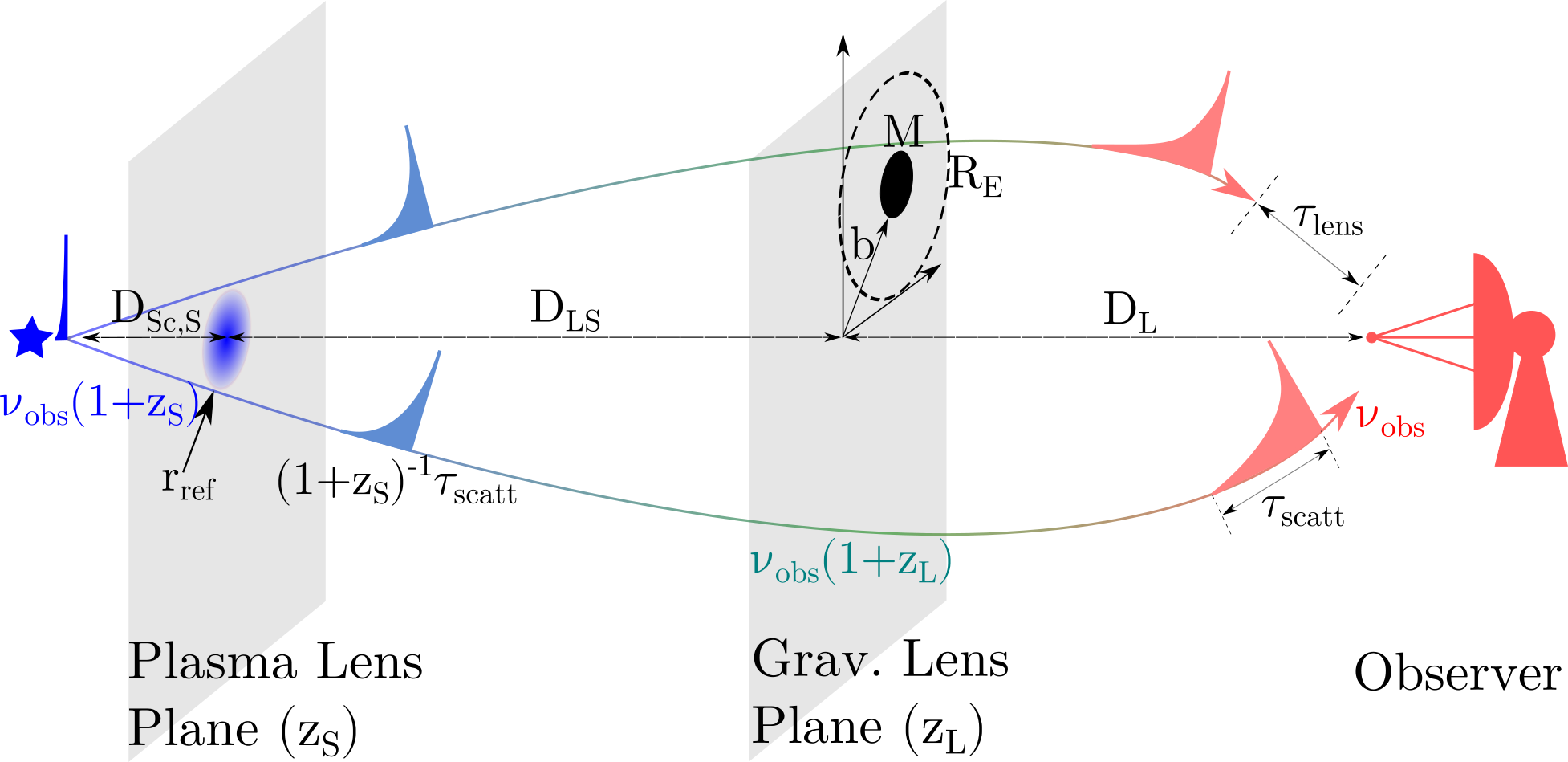}
    \caption{A two-screen model for a coherently lensed FRB observed at some central frequency $\nu_{\mathrm{obs}}$. The plasma lens is responsible for the observed temporal broadening ($\tau_{\mathrm{scatt,obs}}$), produced by a scattering screen of apparent size $r_{\mathrm{ref}}$. The gravitational lens, modeled as a point mass with mass $M$ and impact parameter $b$, can be thought of as a very long baseline interferometer with baseline $\sim R_E \propto \sqrt{M}$ observing at a frequency of
    $\nu_{\mathrm{obs}}(1+z_\mathrm{L})$ from the lens plane. When the scattering screen looks like a point source (Eq.~\ref{eq:unresolved_z}) to the gravitational lens, coherence is maintained, and the observer can see an interference fringe.}
\label{fig:two_screen}
\end{figure*}
\section{Two-Screen Model for FRB Lensing}
\label{sec:screens}
To quantify the effect of scattering from inhomogeneous plasma, we construct a two-screen model involving a gravitational lens and a plasma lens, or scattering screen. We will see that the scattering screen is crucial to this analysis, and in the following subsections we describe how we augment the point-lens model with a scattering screen (see Fig.~\ref{fig:two_screen}). The extent of the scattering depends not on the total density of the plasma but rather on the fluctuations in the density about the mean. The plasma in the Milky Way is known to exhibit a power-law distribution of density fluctuations. The so-called ``big power law in the sky'' spans over 10 orders of magnitude
in length scale~(see e.g.~\citep{armstrong1995electron,lee2019interstellar}), and is responsible for radio propagation effects such as the scattering and scintillation seen in pulsars and other radio transients~\citep{narayan1992physics,macquart2019spectral,schoen2021scintillation}. 

\subsection{Thin Plasma Screen}
The three-dimensional density fluctuations that cause scattering and scintillation are often modeled as a thin screen of two-dimensional electron density inhomogeneities at some effective distance $D_{Sc,S}$ from the source~\citep{lorimer2004handbook}. For instance, in a single-screen model where $D_{Sc,S}$ is the distance from the screen to the source and $D_{Sc,O}$ is the distance from the screen to the observer, the electric field amplitude can be expressed as a Fresnel integral
over the 2D screen plane coordinate $\vec{\rho}$ (see, e.g.,~\citep{feldbrugge2019oscillatory}):
\begin{align}
    E(\nu) = &\int~d^2\vec{\rho} \exp(2\pi i\delta n(\vec\rho) K / \nu) \nonumber\\
    \times &\exp(2\pi i \nu \rho^2 (D_{Sc,S}^{-1} + D_{Sc,O}^{-1}) / 2),\label{eq:fresnel}
\end{align}
where we assume $K = 1/\SI{2.41e-4}{\centi\meter^{-3}~\parsec~ \mega\hertz^{-2} \second^{-1}}$, and represent density fluctuations as an 2D scalar field $\delta n(\vec{\rho})$ (collapsed over the propagation axis); these density fluctuations source phase fluctuations which are amplified as $\nu^{-1}$ at low frequencies. 

It is evident that the effective distance $(D_{Sc,S}^{-1} + D_{Sc,O}^{-1})^{-1}$, is dominated by the smaller of the two distances. Since FRBs originate from dense local environments within their host galaxies~\citep{chawla2021modeling}, the temporal broadening is often assumed to originate in the host environment/galaxy. This is equivalent to taking $D_{\rm eff} = (D_{Sc,S}^{-1} + D_{Sc,O}^{-1})^{-1} \approx D_{Sc,S}$. We use this approximation for the remainder of this work, although we acknowledge that a complete treatment may
need to contend with multiple scattering screens in cases where amount of scattering from the Milky Way and the host galaxy are comparable~\cite{katz2020looking}. %

While the path integral formalism is formally correct, it renders performing accurate calculations difficult~\citep{feldbrugge2019oscillatory}, and there are simpler characterizations that succinctly capture the relevant physics. For example, one can define the distance on the screen over which the RMS phase fluctuation approaches 1 radian. This can be interpreted as the size of a coherent spatial patch on the screen, and is referred to
as $r_{\rm diff}$~\citep{narayan1992physics}. A similar parameterization uses the fact that spatially inhomogeneous phase shifts lead to angular deflections and thus multi-path propagation from the screen to the observer. This leads to temporal pulse broadening over a timescale
\begin{equation}
c\tau_{\rm scatt} \sim \dfrac{r_{\textrm{ref}}^2}{2D_{Sc,S}} + \dfrac{r_{\textrm{ref}}^2}{2D_{Sc,O}}
\label{eq:tau_scatt}
\end{equation}
where $r_{\rm ref} \gg r_{\rm diff}$ is the effective transverse size of the screen~\cite{masui2015dense,katz2020looking}. We see in this picture that the scattering is symmetric under exchanging $D_{Sc,S}$ and $D_{Sc,O}$, and again that is dominated by the shorter of the two path lengths. Furthermore, the two pictures can be related by $r_{\rm diff} r_{\rm ref} = D_{\rm eff} c / (2\pi \nu_{\rm obs})$~\citep{narayan1992physics}, where $D_{\rm eff}^{-1} = D_{Sc,S}^{-1} + D_{Sc,O}^{-1}$. 
This geometric model is a useful simplification because unlike $r_{\rm diff}$ or $r_{\rm ref}$, $\tau_{\rm scatt}$ can be measured for FRBs~\citep{masui2015dense,pleunis2021fast} and allows us to use an observable to constrain the unknown scattering physics. Doing so eliminates one of the two model parameters ($r_{\rm diff}$); the only remaining astrophysical uncertainty associated with scattering is the effective screen distance $D_{\rm eff}$, assumed to be $D_{Sc,S}$. 

Using this picture, we estimate the FRB's transverse size on the screen, and find that it far exceeds the emission region's intrinsic size (Sec.~\ref{sec:source_size} of $\SI{\sim2e10}{\centi\meter}$).
\begin{equation}
    r_{\rm ref} \sim \sqrt{2c\tau_{\rm scatt} D_{Sc,S}} \sim 10^{13} \text{cm} \left( \dfrac{D_{Sc,S}}{\text{pc}} \dfrac{\tau_{\rm scatt}}{\text{ms}} \right)^{1/2}.
    \label{eq:r_ref}
\end{equation}
If the screen (of apparent size $r_{\rm ref}$) is not resolved by the gravitational lens, coherence is preserved (Sec.~\ref{sec:unresolved}). If the screen is too large, it will be resolved by the gravitational lens. This may cause a drop in sensitivity (Sec.~\ref{sec:resolved}).

\subsection{Unresolved Screens}\label{sec:unresolved}
In the limit that the lensing screen is unresolved by the gravitational-lens interferometer, the screen phase is the same in both ``interferometer paths'' along which the light propagates to the observer. In this case, the inhomogeneities on the plasma scattering screen can be arbitrarily strong. The phase $\varphi(\nu)$ imparted to each path can vary rapidly, as long as their difference is less than a radian. In the absence of cosmological redshift effects, the condition for this (see Sec. 3.2 in~\citep{katz2020looking}) is that 
\begin{equation}
    2\pi \nu_{\mathrm{obs}} r_{\rm ref}(D_{Sc,S},\tau_{\mathrm{scatt,obs},j}) R_E / D_{L} < 1.
    \label{eq:unresolved}
\end{equation}
We modify this calculation for the case where the scattering is in the host galaxy $(D_L \to D_{LS})$. We also take into account that the observed pulse broadening timescale $\tau_{\mathrm{scatt,obs},j}$ has experienced time dilation: in the host frame where the scattering occurs, the pulse broadening timescale is $\tau_{\mathrm{scatt,obs},j}(1+z_{\mathrm{S},j})^{-1}$. Finally, we note that the frequency of the radio emission
gets redshifted as it propagates from the source to the observer. The phase is accumulated over the entire path length, but most of the phase difference occurs in the vicinity of the lens; it is a good approximation to replace $\nu_{\mathrm{obs}} \to \nu_{\mathrm{obs}}(1+z_\mathrm{L})$. Eq.~\ref{eq:unresolved} with these modifications becomes 
\begin{equation}
    2\pi \nu_{\mathrm{obs}} r_{\rm ref}(D_{Sc,S},\tau_{\mathrm{scatt,obs},j}) R_E / D_{LS} < \dfrac{\sqrt{1+z_{\mathrm{S},j}}}{1+z_\mathrm{L}}.
    \label{eq:unresolved_z}
\end{equation}

Eq.~\ref{eq:unresolved_z} is satisfied when either the gravitational lens has less resolving power (small $R_E$ or large $\lambda$), or a small scattering screen. The latter can be accomplished either by using bursts with a short scattering timescale or some knowledge of the scattering screen's distance from the source, as shown in Eq.~\ref{eq:r_ref}.  Scattering timescales vary by 2-3 orders of magnitude from burst to burst. Therefore we enforce Eq.~\ref{eq:unresolved_z} for each burst individually using the measured pulse broadening timescale $\tau_{\mathrm{scatt,obs},j}$ and inferred redshift ($z_{S,j}$) for each burst. We conservatively assume that all the broadening originates in the source-local environment (i.e. we do not attempt to subtract off the Milky Way contribution to the pulse broadening). Then, we assume several different values of $D_{Sc,S}$ to calculate the screen
size. In the CHIME/FRB analysis pipeline, $\tau_{\rm scatt,i}$ is currently measured at $\nu = \SI{600}{\mega\hertz}$ and is assumed to scale as $\nu^{-4}$~\citep{lorimer2004handbook}. However, it is currently measured with low-resolution intensity data which cannot resolve scattering timescales $\lesssim \SI{100}{\micro\second}$. As a result, $\approx 30$ of our scattering timescale measurements are upper limits~\citep{kader2022high}. We restrict the $\lambda_{ij}$ integral in Eq.~\ref{eq:n_delta} to regions in the $M-z_L$ plane where Eq.~\ref{eq:unresolved_z} is satisfied. The resulting lensing rate as a function of lens mass, for a variety of screen distances, is plotted in the right panel of Fig.~\ref{fig:optical_depth}. It is evident that scattering screens reduce our sensitivity to large time delays in the presence of scattering screens far away from the source.

\subsection{Screen's Proximity to FRB Source}
By constraining the scattering timescale for each burst, we have translated the astrophysical uncertainties associated with decoherence into a single parameter $D_{\rm eff} \approx D_{Sc,S}$ -- the effective distance between the scattering screen and the FRB. What is a representative median value of $D_{Sc,S}$, averaged over a sample of CHIME-detected FRBs? We estimate this by considering possible origins for the excess scattering in extragalactic FRBs compared to Galactic pulsars, as
established in~\citep{chawla2021modeling}. One explanation is that FRBs are scattered by dense clouds in the circumgalactic medium (CGM) of intervening galaxies~\citep{mccourt2018characteristic,vedantham2019radio,chawla2021modeling}. In this case, the effective distance could be on the order of \SIrange{10}{100}{\mega\parsec}, and Eq.~\ref{eq:r_ref} implies that the source's apparent size ($\sim r_{\rm ref}$) would also be correspondingly large. However, there is growing evidence that the excess scattering is not dominated by clouds in the CGM.

First, if CGM scattering were an explanation for the excess scattering present in the population of FRBs, the screen from the Milky Way would resolve the angular broadening from the CGM, and scintillation would not be regularly observed in FRBs. The observation of both scintillation and scattering in FRB 110523~\citep{masui2015dense} indicates that the angular broadening is unresolved by a Milky-Way screen, and leads to a direct constraint on the scattering geometry for FRB 110523 of $D_{Sc,S} \lesssim 44$ kpc. While this
is only a single example, further examples of FRBs with spectral structure consistent with diffractive scintillation have been identified in ASKAP~\citep{macquart2019spectral}, UTMOST~\citep{farah2018frb}, and CHIME/FRB bursts~\citep{schoen2021scintillation}. 

Second, leading models of CGM scattering~\citep{vedantham2019radio} assume that the CGM efficiently scatters radio waves. This efficiency is quantified through the fluctuation parameter $\widetilde{F}$. Observationally, $\widetilde{F}$ is proportional to the ratio $\tau / \mathrm{DM}^2$, and depends on factors including the filling factor of the gas, the size distribution of cloudlets within the gas, the size of density fluctuations within those cloudlets, and the inner/outer scales
of the turbulence. While the Local Group may not be representative of the CGM of intervening galaxies, measurements of $\tilde{F}$ from the Local Group~\citep{ocker2021constraining} indicate empirically that $\tilde{F}$ is two orders of magnitude smaller  than that assumed by leading theories ($\tilde{F} \sim \SI{500}{\parsec^{-2/3}\kilo\meter^{-1/3}}$,~\citep{vedantham2019radio}) where the CGM provides the observed anomalous scattering~\citep{chawla2021modeling}. This significantly diminishes the possibility that the CGM of intervening galaxies can provide the observed scattering. 

Two remaining possibilities are that the scattering is provided by the host galaxies or local environments of FRBs. Clues about the environments surrounding FRB hosts for individual specimens~\citep{masui2015dense, michilli2018extreme} as well as population studies~\citep{chawla2021modeling} provide evidence that many, if not most, FRBs are associated with special regions of their host galaxies. 
This has been directly confirmed with the VLBI localizations of \revCL{FRB 20121102A}~\citep{marcote2017repeating} to a star-forming region whose H-$\alpha$ radius is $\sim 460$ pc~\citep{kokubo2017halpha}. Similarly, \revCL{FRB 20180916B} is only 250 pc away from (but not residing within) a \SI{1.5}{\kilo\parsec}-long, V-shaped region of star formation in its host galaxy~\citep{marcote2020repeating,tendulkar2021pc}.
The spatial association of a large fraction of precisely localized FRBs~\citep{mannings2021high} with spiral arms in their host galaxies and the random host galaxy inclination angles robustly constrains the value of $D_{Sc,S}$, averaged over all bursts, to be at most the disk thickness of a galaxy. While disk thicknesses may vary from those measured from the Milky Way, the scale height of electron fluctuations in the Milky Way is $\approx
\SI{750}{\parsec}$~\citep{ocker2020electron}, yielding $D_{Sc,S} \lesssim \SI{1}{\kilo\parsec}$. In addition, a detailed population synthesis study
investigating the DM and scattering timescale distributions~\citep{chawla2021modeling} provides tentative evidence for a significant contribution to the scattering from dense, small-scale clumps near the unknown FRB central engine. 

This is a very plausible hypothesis. We note that the Crab Nebula, which dominates the temporal broadening observed in Crab pulses, has a physical extent of $\approx \SI{1.6}{\parsec}$. Detailed studies of Crab scattering reveal plasma screen structures located $\approx \SI{2}{\parsec}$ away from the pulsar~\citep{backer2000plasma}. Pulsars or magnetars with sufficient rotational/magnetic energy (i.e., those that are young enough) to produce bursts luminous enough to be seen at a cosmological distance ($10^5$ times brighter than a fiducial Crab giant pulse; see~\citep{cordes2016supergiant}) could very well be, on average, embedded in similar or even more compact host environments. The dense environment surrounding a decades- or
century-old magnetar which could produce an FRB~\citep{margalit2018concordance} can easily host parsec-scale structures which explain the observed scatter-broadening. In the rest of our analysis, we therefore assume a fiducial screen position of $\sim 1$ pc, though we also quote results for $0.1$,$10$, and $\SI{100}{\parsec}$.

\subsection{Resolved Screens}\label{sec:resolved}
Until now we have only considered detecting coherent FRB lensing in systems where the scattering screen is unresolved. If the screen is resolved, our signal, which is a peak in the measured time-lag autocorrelation function of an FRB, may be washed out. This is similar in spirit to femtolensing PBH constraints; however, we emphasize that a point source behind a scattering screen is not equivalent to a \textit{bona fide} incoherent extended source. The former has a characteristic decorrelation bandwidth ($\Delta\nu_{\mathrm{bw}}$) over which the signal remains coherent, whereas the latter does not.

For this reason the signal can be recovered in certain circumstances. Since we search for peaks in the ACF, we are most sensitive when peaks are localized to one delay bin ($\delta \tau < 1.25$ ns). If the screen is resolved by the lens the peak may be washed out among many ($\sim \tau_{\rm scatt} / 1.25 \text{~ns}$) trial delays. This
reduces sensitivity in a search based on finding peaks in the ACF. Another way of saying this is that each subband in the dynamic spectrum of bandwidth $\nu_{dec}$ (the decorrelation bandwidth of the FRB emission) will have a different observed delay which is a sum of the gravitational lensing delay and a frequency-dependent scattering screen delay. For voltage data covering a bandwidth of
$\nu_{\mathrm{bw}}$, the height of the lensing peak at $\tau_{\mathrm{lens}}$ would be suppressed by a factor of $\Delta \nu_{\mathrm{dec}}/\Delta \nu_{\mathrm{bw}}$. While our search has a wide bandwidth $\Delta\nu_{\mathrm{bw}} = \SI{400}{\mega\hertz}$, it may the case that e.g., in a sub-banded search, $\Delta\nu_{dec} \approx \Delta\nu_{bw}$. The coherence would be maintained within each sub-band and sub-bands could then be incoherently combined. At present, however, we do not attempt to detect resolved coherent lensing, so we do not include this regime in our present constraints. We defer working in the limit of a resolved screen to future work.

There is another way to circumvent the decoherence inflicted by scattering screens. Decorrelation bandwidths (equivalently, the scattering timescales) are highly frequency-dependent, and searches for FRBs lensing at higher frequencies may exploit the steep frequency scaling of scattering timescales to leverage this. Hence, it is very possible that even if the screen is resolved at low frequencies, coherence is maintained at higher frequencies. FRBs have been detected at frequencies down to
\SI{110}{\mega\hertz}~\citep{chawla2020detection,pleunis2021lofar,pastor2021chromatic} and up to 8 GHz~\citep{michilli2018extreme}, making this a promising and straightforward possibility.

\section{Constraints}
Having calculated the expected lensing rate $\lambda(M_c)$ first without and now, with the effect of scattering screens (left and right panels of Fig.~\ref{fig:optical_depth}), we aim to translate those rates into constraints on the PBH abundance. First, we note that as a result of Eq.~\ref{eq:mass_function}, all of our measurements of the lensing rate have a simple linear dependence on $f(M_c)$, the fraction of the cosmological cold dark matter density ($\Omega_{\rm c} \rho_{crit}$) that is composed of PBHs of mass $M_c$, assuming a mass function peaked around $M_c$. It is convenient to define the function $\lambda_1(M_c)$ as the lensing rate assuming $f(M_c)=1$; from here on we can write the actual lensing rate as $\lambda = f \lambda_1$. In the remainder of this section, we will omit the $M_c$ arguments for brevity, though $f$,$\lambda$, and $\lambda_1$ are functions of $M_c$.

The exclusion limit can be thought of as an estimator $\hat f$ of the true value $f$ satisfying $\hat f > f$ with high probability. However, the estimator depends on the mass function assumed. In our case, the mass function is parameterized by a single parameter $M_c$, so $\hat{f}$, like $f$, is a function of $M_c$.
To constrain compact dark matter, we may employ either a frequentist or a Bayesian framework, which have different formalisms for calculating $\hat{f}$. The process of detecting lenses can be modeled as a Poisson process with rate $\lambda = f \lambda_1$. Then the probability of observing $k$ lensing events is 
\begin{equation}
    P(k|\lambda) = e^{-\lambda} \lambda^k / k!. 
    \label{eq:poisson}
\end{equation}

In a frequentist framework~\citep{niikura2019microlensing}, the probability of getting our null search result ($k = 0$ coherent lensing events in our sample) is
\begin{equation}
    P(k=0|\lambda) = \exp(-\hat{f}  \lambda_1(M_c)).
    \label{eq:frequentist_zero}
\end{equation}
The inequality $P(k=0|\lambda) < 0.05$ constrains our false non-detection rate and solving it sets $\hat{f}$. If, for example, we made either $k = 0$ or $k = 1$ detections in our entire search, the left side of Eq.~\ref{eq:frequentist_zero} would instead be $P(k=0|\lambda) + P(k=1|\lambda)$. We would instead solve the following inequality for $\hat{f} $: 
\begin{equation}
    \exp(-\hat{f}  \lambda_1)[1 + \hat f \lambda_1 ] < 0.05.
    \label{eq:frequentist_one}
\end{equation}

\begin{figure*}
    \centering
    \includegraphics[width = 0.9\textwidth]{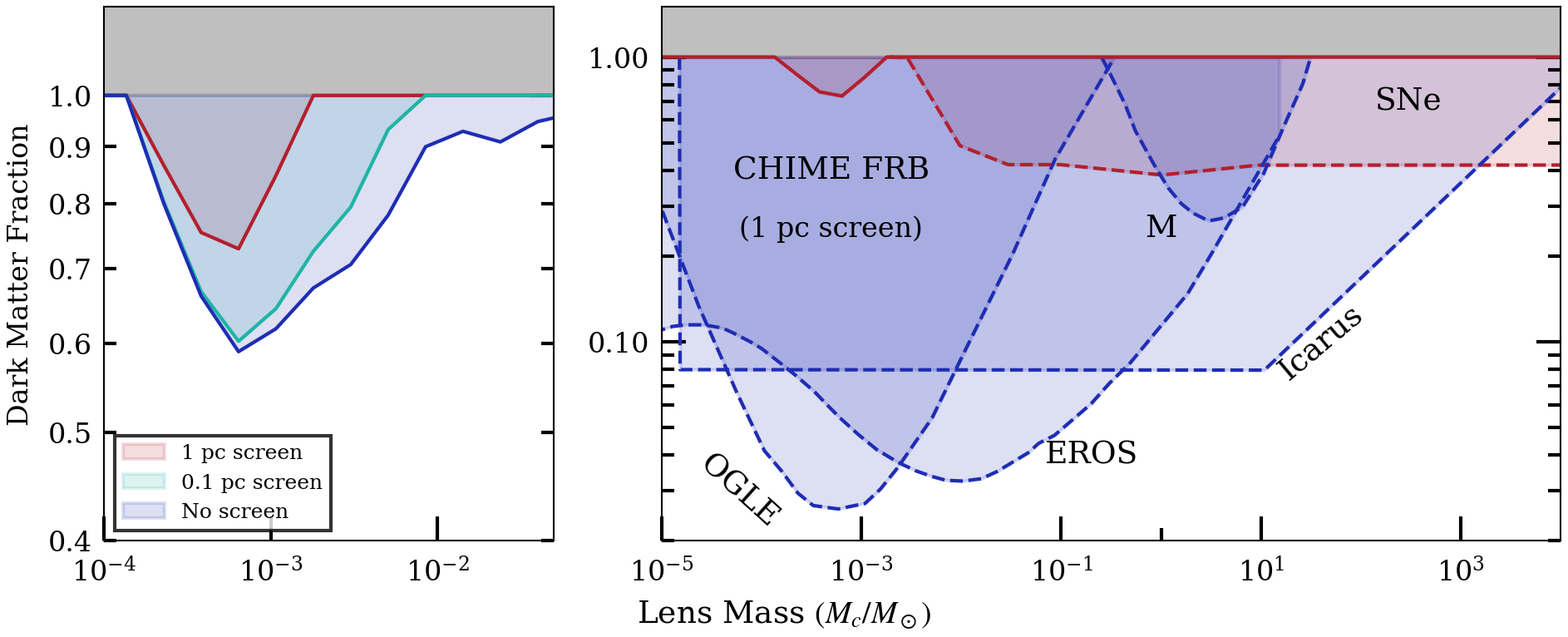}
    \caption{Left: 95\% Constraints on PBHs as a function of scattering screen distance corresponding to the optical depth calculated in Fig.~\ref{fig:optical_depth}. We plot our fiducial (1 pc screen) model in red and suppress curves for screen distances of 10 and 100 pc because $\lambda < 3$ under those assumptions. Right: A collection of microlensing constraints on the fraction of dark matter composed of compact objects (such as PBHs), $f(M_c)$, assuming a monochromatic mass function peaked around $M_c$. We have shown Local Group PBH constraints in blue (M: MACHO~\citep{alcock2001macho}, EROS~\citep{tisserand2007limits}, OGLE~\citep{niikura2019microlensing}, Icarus~\citep{oguri2018understanding}), and Local Universe constraints in red
    (SNe~\citep{zumalacarregui2018limits}, CHIME/FRB, this work). CHIME/FRB lensing constraints depend on our two-screen scattering model, in which we have assumed that the average FRB is scattered by a screen at an effective distance of $\SI{1}{\parsec}$, and our model for how DM correlates with distance. In these constraints, we have used Eq.~\ref{eq:frequentist_zero} to define the exclusion limit as a function of $M_c$. Wave optics effects suppress our signal at $M \lesssim 1.5\times 10^{-4} M_{\odot}$ and finite source size suppresses our signal at $M \gtrsim 3\times 10^4 M_{\odot}$.
    This shows that coherent FRB lensing has the potential to search new parameter space for exotic compact objects such as PBHs.}
    \label{fig:exclusion}
\end{figure*}

In a Bayesian framework, the excluded region is defined instead by the following condition:
\begin{equation}
    \label{eq:bayes}
    0.05 > p(f > \hat{f}  | k=0)
\end{equation}
where $k=0$ denotes non-detection of lensing. Informally Eq.~\ref{eq:bayes} can be thought of as the probability of being ``wrong'' about $\hat f$, requiring that the true value of $f$ has only a $5\%$ chance of being higher than our inferred value $\hat{f} $. We expand the right hand side of Eq.~\ref{eq:bayes} in terms of the posterior  $p(f|k=0)$:
\begin{align*}
    p(f &> \hat{f}  | k=0) = \\
     &\int_{\hat f}^\infty p(f|k=0)~df .
    \intertext{which in turn can be re-written using Bayes's theorem:}
    p(f|k=0) &\propto p(k=0|f) p(f) . \\
    \intertext{The first factor on the right-hand side is simply the Poisson likelihood}
    p(k=0|f) &= \exp(- f \lambda_1),\\
    \intertext{and the second term is a prior on $f$ which we take to be the uniform distribution supported from $f = 0$ to some cutoff value $F$.} 
    \intertext{If we normalize $p(f|k=0)$ we obtain}
    p(f|k=0) &= \dfrac{e^{- f \lambda_1}}{1 - e^{- F \lambda_1}}
    \intertext{which we substitute back into Eq.~\ref{eq:bayes} to yield the Bayesian criteria for the excluded region $\hat f$:}
\end{align*}
\begin{equation}
    0.05 = \dfrac{e^{-\hat f \lambda_1} - e^{-F \lambda_1}}{1 - e^{-F \lambda_1}} .
    \label{eq:bayes_zero}
\end{equation}
Both criteria (Eq.~\ref{eq:frequentist_zero},~\ref{eq:bayes_zero}) are valid formulations, and they agree in the limit that $F \to \infty$. We use the frequentist method because it is the most conservative and for consistency with existing constraints~\citep{niikura2019microlensing}. Our final constraints are plotted in Fig.~\ref{fig:exclusion}.

Our constraints are complementary to existing microlensing constraints on compact dark matter~\citep{carr2020primordial}. Conventional microlensing constraints (e.g.,~\citep{niikura2019microlensing}) extend further \revCL{in mass} than the ones described here. However, those constraints are from observations of M31 and thus are only sensitive to compact objects in the local universe. In contrast, our long sightlines extend out to non-negligible redshifts
where the dark matter density is expected to approach its cosmological average value. The constraints most similar to ours in the literature are those from Type Ia supernovae microlensing~\citep{zumalacarregui2018limits}. However, due to our coherent search method, we are sensitive to a lighter mass range with sensitivity approaching that of~\citep{zumalacarregui2018limits}. We expect that with a larger sample of bursts
from CHIME/FRB, our method will soon provide an
independent probe with sensitivity to extragalactic compact dark matter at masses inaccessible by other means.

\section{Discussion and Conclusions}\label{sec:discussion}
There are several limitations to the constraints presented here. First, they are subject to the unknown uncertainties in inferring the redshift of an FRB from its DM, where a large uncertainty arises from assuming a value of $\mathrm{DM_{host}}$. Several clues point toward a wide range of $\mathrm{DM_{host}}$ in the FRB population. First, the recent VLBI localization and host identification of FRB 20190520B has revealed its extremely large $\mathrm{DM_{host}}$ contribution of $\approx \SI{900}{\parsec~\centi\meter^{-3}}$~\citep{niu2021repeating}. This means that for some FRBs, the distance determination in Eq.~\ref{eq:dm_eg} is unreliable. Independently, a statistically-significant spatial correlation has been detected between $z \sim 0.4$ galaxies in large optical surveys and FRBs whose extragalactic DM $\gtrsim\SI{785}{\parsec~\centi\meter^{-3}}$. This can be interpreted as evidence that an order-one fraction of such high-DM FRBs in the CHIME/FRB Catalog have host DMs of $\sim \SI{400}{\parsec~\centi\meter^{-3}}$~\citep{rafieiravandi2021catalog}. 

In light of these two lines of evidence, we have tested two extreme scenarios to estimate the uncertainty on our constraints resulting from distance determination. In one scenario, we assume that all three bursts in our sample with large total DM ($>
\SI{1000}{\parsec~\centi\meter^{-3}}$) are similar to 20190520B. This is conservative because to the best of our knowledge, FRBs like 20190520B are not representative of the population of FRBs detected by CHIME/FRB. Like FRB 20121102A, the properties of FRB 20190520B (e.g. rotation measure and host galaxy) are quite different from other FRBs localized by ASKAP~\citep{macquart2020census,bhandari2022characterizing,niu2021repeating}. In this scenario, the total lensing rate shown in Fig.~\ref{fig:optical_depth} is reduced by $\approx 15\%$. In another scenario, Ref.~\citep{rafieiravandi2021catalog} implies that some high-DM FRBs (DM $\gtrsim$\SI{785}{\parsec~\centi\meter^{-3}}) have a large host contribution. To conservatively model this scenario, we double the $\mathrm{DM_{host}}$ of all bursts in our sample with $\mathrm{DM} > \SI{500}{\parsec~ \centi\meter^{-3}}$ FRBs from 117 to \SI{234}{\parsec~\centi\meter^{-3}}. In this scenario, the optical depth is reduced by $\approx 20\%$. 

These two scenarios bracket the uncertainty in our optical depth arising from DM-based distances. We emphasize that this uncertainty is a short-term problem motivating a long-term solution: to localize and follow up FRBs using upcoming instruments like CHIME/FRB Outriggers~\citep{leung2020synoptic,cassanelli2021localizing,mena2022clock,cassanelli2022fast} to directly obtain their host galaxies' redshifts. 

Second, our constraints are sensitive to the measured scattering timescale of each burst, which we use to estimate the extent of the plasma decoherence. At present, the most mature CHIME/FRB pipeline for measuring burst scattering timescales (intensity \texttt{fitburst}) uses low-resolution ``intensity'' data~\citep{FRBSystemOverview}. The intensity data's time resolution limits its ability to measure scattering tails with 
timescales \revCL{shorter} than $\approx \SI{100}{\us}$. For bursts
where a scattering tail is not detected, an upper limit on the scattering of $\SI{100}{\us}$ is adopted. This is a very conservative treatment given that nanosecond timescales have been observed in at least one FRB~\citep{bhardwaj2021nearby,kirsten2021repeating,nimmo2021burst}. In the future, adapting the \texttt{fitburst} analysis pipeline (described in detail in~\citep{masui2015dense,chimefrbcatalog1}) to use CHIME/FRB baseband data would allow for a higher time resolution of $\approx
\SI{2.56}{\us}$. More accurate estimates of the screen size will lead to improved constraints. 

Third, our search method assumes that the scattering screen is unresolved by the gravitational lens, and is insensitive to screens resolved by the lens. Our calculations demonstrate that this region of parameter space corresponds to solar-mass lenses, a mass range that has enjoyed renewed interest due to the detection of gravitational waves from compact binary mergers~\citep{abbott2016observation}. One way to access this mass range with FRB gravitational-lens interferometry is by developing more sophisticated correlation algorithms to extract a lensing signal from the data with knowledge of the properties of the scattering screen, which may be measured from the data themselves. Another is to change the observing frequency: though finite source size is less of a hindrance at CHIME frequencies, both wave optics effects and scattering screens are less of a problem at higher frequencies. This will broaden the reach of a search at the low-mass end ($\propto \nu^{-1}$ due to reduced wave effects) and at the high-mass end (smaller $r_{\mathrm{ref}}$) because scattering timescales are extremely frequency-dependent $(\propto \nu^{-4})$.

The future of FRB gravitational-lens interferometry is bright. Over 3000 FRBs have been detected by CHIME; these will enable the expansion of this sample by over an order of magnitude. In the future, FRBs will be routinely localized; this will provide robust distance measurements. Upcoming FRB surveys with localization capabilities such as the Canadian Hydrogen Observatory and Radio-transient Detector (CHORD)~\citep{vanderlinde2019canadian} and the Deep Synoptic Array (DSA-2000)~\citep{hallinan2019dsa}, will detect FRBs at an even higher rate, and access frequencies up to $\SI{1.5}{\giga\hertz}$ to better exploit
the favorable scalings of coherent FRB lensing at higher frequencies.

In conclusion, this work and its companion paper~\citep{kader2022high} have demonstrated a novel method of searching for coherently lensed FRBs, and have demonstrated the ability of coherent FRB lensing to constrain the constituents of the cosmological dark matter, e.g., primordial black holes. We have quantified the amount of decoherence using a two-screen model containing a gravitational lens plane and a plasma screen; we find that the degree of decoherence is sensitive to the plasma screen geometry. The reach of coherent FRB lensing is increased as the FRB looks increasingly like a point source as viewed from the lens plane. This is quite possible: in some cases FRB emission only involves small amounts of temporal broadening~\citep{nimmo2021burst}. In other cases, studies of FRB properties~\citep{masui2015dense,michilli2018extreme}, their host
environments~\citep{tendulkar2021pc,mannings2021high} and population
studies~\citep{chawla2021modeling,bhandari2022characterizing} of large samples of FRBs support progenitor theories involving young neutron stars~\citep{cordes2016supergiant,margalit2018concordance} with atypical scattering environments (i.e. more extreme and compact than those of pulsars). Like in the case of GRB femtolensing~\citep{katz2018femtolensing}, the finite angular size of the FRB emission region imposes a fundamental high mass sensitivity cutoff for coherent FRB lensing. On the
opposite end of the mass range, wave optics effects complicate searches for compact objects whose Schwarzschild radii are smaller than the wavelength of light used~\citep{katz2020looking}. Despite these limitations, and the uncertainties due to scattering screens, our present results establish the
sensitivity of coherent FRB lensing as a probe of sub-solar mass primordial black holes. They also strongly suggest that it is promising to conduct future searches for coherent lensing at higher observing frequencies, where scattering and wave optics effects are reduced. More broadly, this work establishes the viability of using coherent FRB lensing as a unique tool with broader applications in astrophysics and cosmology.

\begin{acknowledgments}
We acknowledge that CHIME is located on the traditional, ancestral, and unceded territory of the Syilx/Okanagan people. We are grateful to the staff of the Dominion Radio Astrophysical Observatory, which is operated by the National Research Council of Canada.  CHIME is funded by a grant from the Canada Foundation for Innovation (CFI) 2012 Leading Edge Fund (Project 31170) and by contributions from the provinces of British Columbia, Qu\'{e}bec and Ontario. The CHIME/FRB Project is funded by a grant from the CFI 2015 Innovation Fund (Project 33213) and by contributions from the provinces of British Columbia and Qu\'{e}bec, and by the Dunlap Institute for Astronomy and Astrophysics at the University of Toronto. Additional support was provided by the Canadian Institute for Advanced Research (CIFAR), McGill University and the McGill Space Institute thanks to the Trottier Family Foundation, and the University of British Columbia.
\allacks
\end{acknowledgments}
\appendix
\bibliography{radio,stronglensing,chimefrbpapers}


\end{document}